\documentstyle[12pt]{article}

\catcode`\@=11
\long\def\@makefntext#1{
\protect\noindent \hbox to 3.2pt {\hskip-.9pt
$^{{\ninerm\@thefnmark}}$\hfil}#1\hfill}		

\def\@makefnmark{\hbox to 0pt{$^{\@thefnmark}$\hss}}  

\def\ps@myheadings{\let\@mkboth\@gobbletwo
\def\@oddhead{\hbox{}
\rightmark\hfil\ninerm\thepage}
\def\@oddfoot{}\def\@evenhead{\ninerm\thepage\hfil
\leftmark\hbox{}}\def\@evenfoot{}
\def\sectionmark##1{}\def\subsectionmark##1{}}

\renewcommand{\thefootnote}{\fnsymbol{footnote}}

\newcounter{sectionc}\newcounter{subsectionc}\newcounter{subsubsectionc}
\renewcommand{\section}[1] {\vspace*{0.6cm}\addtocounter{sectionc}{1}
\setcounter{subsectionc}{0}\setcounter{subsubsectionc}{0}\noindent
	{\normalsize\bf\thesectionc. #1}\par\vspace*{0.4cm}}
\renewcommand{\subsection}[1] {\vspace*{0.6cm}\addtocounter{subsectionc}{1}
	\setcounter{subsubsectionc}{0}\noindent
	{\normalsize\it\thesectionc.\thesubsectionc. #1}\par\vspace*{0.4cm}}
\renewcommand{\subsubsection}[1]
{\vspace*{0.6cm}\addtocounter{subsubsectionc}{1}
	\noindent {\normalsize\rm\thesectionc.\thesubsectionc.\thesubsubsectionc.
	#1}\par\vspace*{0.4cm}}

\newcounter{appendixc}
\newcounter{subappendixc}[appendixc]
\newcounter{subsubappendixc}[subappendixc]

\renewcommand{\appendix}[1] {\vspace*{0.6cm}
        \refstepcounter{appendixc}
        \setcounter{figure}{0}
        \setcounter{table}{0}
        \setcounter{equation}{0}
        \renewcommand{\thefigure}{\Alph{appendixc}.\arabic{figure}}
        \renewcommand{\thetable}{\Alph{appendixc}.\arabic{table}}
        \renewcommand{\theappendixc}{\Alph{appendixc}}
        \renewcommand{\theequation}{\Alph{appendixc}.\arabic{equation}}
        \noindent{\bf Appendix \theappendixc #1}\par\vspace*{0.4cm}}

\def\abstracts#1{{
	\centering{\begin{minipage}{12.2truecm}\footnotesize\baselineskip=18pt\noindent
	\centerline{\footnotesize ABSTRACT}\vspace*{0.3cm}
	\parindent=0pt #1
	\end{minipage}}\par}}

\renewenvironment{thebibliography}[1]
	{\begin{list}{\arabic{enumi}.}
	{\usecounter{enumi}\setlength{\parsep}{0pt}
\setlength{\leftmargin 1.25cm}{\rightmargin 0pt}
	 \setlength{\itemsep}{0pt} \settowidth
	{\labelwidth}{#1.}\sloppy}}{\end{list}}

\newcounter{itemlistc}
\newcounter{romanlistc}
\newcounter{alphlistc}
\newcounter{arabiclistc}

\newcommand{\fcaption}[1]{
        \refstepcounter{figure}
        \setbox\@tempboxa = \hbox{\footnotesize Fig.~\thefigure. #1}
        \ifdim \wd\@tempboxa > 6in
           {\begin{center}
        \parbox{6in}{\footnotesize\baselineskip=20pt Fig.~\thefigure. #1}
            \end{center}}
        \else
             {\begin{center}
             {\footnotesize Fig.~\thefigure. #1}
              \end{center}}
        \fi}

\newcommand{\tcaption}[1]{
        \refstepcounter{table}
        \setbox\@tempboxa = \hbox{\footnotesize Table~\thetable. #1}
        \ifdim \wd\@tempboxa > 6in
           {\begin{center}
        \parbox{6in}{\footnotesize\baselineskip=20pt Table~\thetable. #1}
            \end{center}}
        \else
             {\begin{center}
             {\footnotesize Table~\thetable. #1}
              \end{center}}
        \fi}

\def\@citex[#1]#2{\if@filesw\immediate\write\@auxout
	{\string\citation{#2}}\fi
\def\@citea{}\@cite{\@for\@citeb:=#2\do
	{\@citea\def\@citea{,}\@ifundefined
	{b@\@citeb}{{\bf ?}\@warning
	{Citation `\@citeb' on page \thepage \space undefined}}
	{\csname b@\@citeb\endcsname}}}{#1}}

\newif\if@cghi
\def\cite{\@cghitrue\@ifnextchar [{\@tempswatrue
	\@citex}{\@tempswafalse\@citex[]}}
\def\citelow{\@cghifalse\@ifnextchar [{\@tempswatrue
	\@citex}{\@tempswafalse\@citex[]}}
\def\@cite#1#2{{$\null^{#1}$\if@tempswa\typeout
	{IJCGA warning: optional citation argument
	ignored: `#2'} \fi}}

 1
 1
 1

\font\ninerm=cmr9

\begin{document}

\centerline{\normalsize\bf }
\baselineskip=22pt
\centerline{\normalsize\bf Neutron pick-up pion transfer reactions}
\centerline{\normalsize\bf for}
\centerline{\normalsize\bf the formation of deeply bound pionic atoms in 
$^{208}$Pb}
\baselineskip=16pt
\centerline{\normalsize\bf }

\centerline{\footnotesize S. Hirenzaki$^{a)}$ and H. 
Toki$^{b)}$}
\baselineskip=13pt
\centerline{\normalsize\bf }

\centerline{\footnotesize\it a) Department of Physics, Nara Women's 
Univ., Nara 630, Japan}
\centerline{\footnotesize\it b) Research Center for Nuclear Physics, 
Osaka Univ., Ibaraki, Osaka 567, 
Japan}
\baselineskip=12pt
\vspace*{0.3cm}

\vspace*{0.9cm}
\abstracts{We study $(n,d)$ 
and (d,$^3$He) reactions on $^{208}Pb$ leading to deeply bound 
pionic atoms with one neutron hole being left.  We develop a theoretical 
model to calculate cross sections of deeply bound pionic 
atoms and quasi-elastic pion production. 
The results are compared with data which have 
been obtained so far.  
We show that the theoretical model describes the data of 
both reactions well.  }

\normalsize\baselineskip=20pt
\setcounter{footnote}{0}
\renewcommand{\thefootnote}{\alph{footnote}}
\section{Introduction}
Since the suggestion of Toki and Yamazaki for the formation of deeply 
bound pionic atoms such as $1s$ and $2p$ states in heavy nuclei using direct 
reactions,\cite{Toki88,Toki89} there have been a number of 
experimental attempts to find these states.  These states have a pionic 
halo structure due to a strong repulsive pion-nucleus interaction, which 
saves a pion from being absorbed by the nucleus.\cite{Toki88,Toki89}  
The $(n,p)$ pion transfer reactions on $^{208}Pb$ 
at $T_n = 420 MeV$ were measured by Iwasaki $et$ $al.$ using the Charge 
Exchange facility at TRIUMF.\cite{Iwasaki91}  No evidence for pionic 
atoms was found, the reason for which was interpreted as due to large 
momentum mismatch at the nuclear surface where the pion transfer reaction 
takes place.  A similar conclusion was made for (d,$^2$He) reactions 
performed at SATURNE.\cite{Hayano91}  It was then concluded that the 
pion transfer reactions of this type were not suited for the formation 
of deeply bound 
pionic atoms.\cite{Iwasaki91}  The same conclusion was obtained also by 
Nieves and Oset.\cite{Nieves90}

Different reactions were studied theoretically by other authors.  Nieves and 
Oset studied the reactions induced by leptons such as $(\gamma, \pi^+)$ 
\cite{Nieves90-2} and $(e,e')$.\cite{Nieves91}  They studied also $(\Sigma, 
\Lambda)$ reactions in order to reduce the momentum mismatch occurring in 
$(n,p)$ reactions by reducing the Q-value due to the $\Sigma-\Lambda$ 
mass difference.\cite{Nieves92}  Their latest suggestion on the use of 
low energy in-flight pion is also worth-mentioning, in which 
a photon after the direct capture of the in-flight $\pi^-$ in deeply bound 
pionic states can be observed.\cite{Nieves92-2}

Toki $et$ $al.$, on the other hand, pushed further the use of the neutron 
and deuteron beams, but added another nucleon to participate in the pion 
transfer process; pion transfer followed by one nucleon pick up reactions 
such as 
$(n,d)$ and (d,$^3$He).\cite{Toki91,Hirenzaki91}  
We show the schematic diagram for (d,$^3$He) reaction leading to pionic 
bound states in Fig. 1.  
These processes have an advantage to be able to choose the incident 
energy so as to make the reaction recoilless.  In this kinematics, the 
pion carries a small momentum into the target nuclear system.  
Although the $(n,d)$ and (d,$^3$He) reactions lead 
to complicated nuclear configurations together with the formation of 
pionic atoms, the recoilless condition may make the process 
experimentally feasible.  In fact, the cross sections leading to pionic 
states $(n,l)$ together with a neutron hole states of the same orbital 
angular momentum ('quasi' substitutional configurations, $\Delta l = 0$) 
were shown to be enhanced.\cite{Toki91,Hirenzaki91}.

The first attempt with the use of $(n,d)$ reactions was performed by Trudel 
$et$ $al.$ with the Chargex facility at TRIUMF.\cite{Trudel91}  The zero 
degree experimental data was reported by Yamazaki at the International 
Nuclear Physics Conference (INPC).\cite{Yamazaki93}  The forward double 
differential cross section for the $^{208}Pb(n,d)$ reaction at $T_n = 400 
MeV$ was measured with the experimental energy resolution of about $1 
MeV$. In this experiment we observe that the cross section begins to increase 
gradually from flat background at the excitation energy 
corresponding to about $10 MeV$ below the pion production threshold, 
not at the pion production threshold.  This strength below the 
pion production threshold was expected to come from the deeply bound pionic 
atom formation.  
Matsuoka $et$ $al$. measured cross sections of $(p,pp)$ reactions to 
observe the deeply bound pionic atoms and obtained the similar spectrum 
to that of the $(n,d)$ reactions.\cite{Matsuoka95}  
Thus, following the theoretical predictions,
\cite{Hirenzaki91} the 
experimental search for deeply bound pionic atoms by (d,$^3$He) reactions 
was performed 
in order to improve the data quality using the 
excellent deuteron beam.  Yamazaki $et$ $al.$ performed an experiment 
very recently at GSI and reported that they found a clear peak structure due to 
deeply bound pionic atom formation.\cite{Yamazaki96}  The agreement 
between the data and the theoretical prediction was found very good and this 
fact demonstrated that the theoretical model 
describes the reaction very 
well.  

In this paper we study the one nucleon pick-up reaction leading to deeply 
bound pionic atoms theoretically.  The calculated results are compared 
with data obtained recently.  
In section 2, we describe the formalism 
of calculating the $(n,d)$ and (d,$^3$He) cross sections leading to 
pionic atoms and the pionic continuum states within the effective number 
approach.  In section 3, we show general features of the one nucleon 
pick-up pion transfer reactions by considering $(n,d)$ reactions.  
We present the calculated results of $(n,d)$ 
reactions and compare them with the data.  The (d,$^3$He) reaction is studied in 
section 4.  Section 5 is devoted for summary and discussions.  

\section{Theoretical Model}

In this section, we describe our theoretical model which was developed to 
calculate the $(n,d)$ and (d,$^3$He) reactions leading to deeply bound 
pionic atoms.  
Since we are interested in the energy spectra only around the pion 
production threshold, we postulate that the spectrum depends on the 
energy only through three pionic processes; (1) bound $\pi^-$ production, 
(2) quasi-elastic $\pi^-$ production, and (3) quasi-elastic $\pi^0$ 
production.  There is a neutron hole for $\pi^-$ production processes and 
a proton hole for a $\pi^0$ production process in the final state.  
Other processes are assumed to be independent of the energy 
in this narrow region.  
 In the previous publications, we have developed the effective number 
approach to calculate the bound $\pi^-$ production process.
\cite{Toki91,Hirenzaki91}  In this section we describe our model 
including quasi-elastic processes.  

The $(n,d)$ cross sections for the 
nuclear target are written in the effective number approach as

\begin{equation}
 \left( \frac{d^2 \sigma}{d \Omega dE} \right)_{nA \rightarrow d(A-1) \pi} =
 \left( \frac{d \sigma}{d \Omega } \right)_{nn \rightarrow d \pi}^{lab}
 \sum_{l_\pi, j_n, J} \left( \frac{\Gamma}{2\pi} \frac{N_{eff}}{ (\Delta E)^2 
 +\Gamma^2 /4} + \frac{2}{\pi} E_\pi p_\pi N_{eff} \right)
\end{equation}

\noindent
with 

\begin{equation}
N_{eff} = \frac{1}{2} \sum_{M,m_s} \left| \int \chi^*_f (\vec{r}) 
\xi^*_{\frac{1}{2},m_s}(\sigma) [ \phi^*_{l_\pi} (\vec{r}) \otimes 
\psi _{j_n} (\vec{r},\sigma) ]_{JM} \chi_i(\vec{r}) d^3r d\sigma \right| ^2
\end{equation}
 
\noindent 
The first term in Eq. (1) corresponds to the formation of 
pion bound states and the second 
one to unbound states.  The $\Delta E$ in the first term 
is defined as $\Delta E = Q + m_{\pi} - B_\pi + S_n - 3.519 MeV$ for the 
$(n,d)$ reactions with the 
pion binding energy ($B_\pi$), the neutron separation energy ($S_n$), and 
the reaction Q-value so as to 
make a peak structure at a resonance energy of each 
configuration of pion-particle neutron-hole state.\cite{Toki91} 
For the (d,$^3$He) reactions we need to use $6.787 MeV$ instead of $3.519 
MeV$ that appeared above.\cite{Hirenzaki91}
The $\Gamma$ denotes the width of the bound pionic 
state. The pion binding energies and widths are calculated using the realistic optical 
potential \cite{Toki89}. To calculate neutron separation energies for deep 
neutron states we use the 
potential parameters determined by Speth $et$ $al.$ \cite{Speth77}.  
We use experimental values for $S_n$ when they are available.  
The $E_\pi$ in the second term 
is the energy of unbound pions in the laboratory frame 
defined as 
$E_\pi = - Q - S_n + 3.519 MeV$ for $\pi^-$ production 
in $(n,d)$ reactions and $p_\pi$ is the 
corresponding pion momentum.  The factor $ \frac{2}{\pi} E_\pi p_\pi $ is 
due to the phase volume of the unbound pion.  The quasi elastic $\pi^0$ 
production is included in a similar way.  
The elementary cross section $(d\sigma/d\Omega)_{nn\rightarrow d\pi}$ is 
derived from the $pp \rightarrow d\pi^+$ cross sections by using the charge 
conjugation. In the (d,$^3$He) reactions, we need to use 
$(d\sigma/d\Omega)_{dp\rightarrow t\pi^+}$ as the elementary cross 
section. We show the elementary cross sections in Fig. 2.\cite{Toki91} 

The $\phi_{l_\pi}$ and $\psi _{j_n}$ in $N_{eff}$ are the pion 
and the neutron hole wave functions with a resultant angular momentum 
$J$.  The pion wave function is calculated by solving the Klein-Gordon 
equation with the realistic potential \cite{Toki89}.  We use the 
harmonic oscillator wave function for the neutron wave function for 
simplicity.  The spin wave function $\xi_{\frac{1}{2},m_s}$ with averaging over 
$m_s$ is to take care of the possible spin directions of the neutrons in 
the target.  The $\chi$'s denote the initial and the final distorted 
waves of the projectile and the ejectile, which are treated in the eikonal 
approximation.\cite{Toki91}  

The treatment of deeply bound pionic states is described in detail in 
previous publication.\cite{Toki91,Hirenzaki91} We describe here the 
quasi-elastic pion production part.  The scattering pionic states are 
obtained by solving the Klein-Gordon equation with the finite Coulomb
potential and the optical potential $V_{opt}$, which is written as 

\begin{displaymath}
2 \bar{\omega} V_{opt} = 
- 4 \pi [ b(r) + B(r)] + 4 \pi \vec{\nabla}
\cdot \{ L(r) [ c(r) + C(r) ] \} \vec{\nabla}
\end{displaymath}

\begin{equation}
- 4\pi \{ \frac{p_1  -1 }{2} \nabla^2 c(r) + \frac{p_2 -1}{2} 
\nabla^2 C(r) \}                   ,
\end{equation}

\noindent
where

\begin{equation}
b(r) = p_1 [ \bar{b_0} \rho (r) - \varepsilon_{\pi} b_1 ( \rho_n (r) - 
\rho_p(r))]  ,
\end{equation}
 
\begin{equation}
\bar{b_0} = b_0 - \frac{3k_F}{2 \pi} (b_0^2 + 2 b_1^2 ) , 
\end{equation}

\begin{equation}
c(r) = p_1^{-1} [c_0 \rho (r) - \varepsilon_{\pi} c_1 (\rho_n (r) - \rho_p (r)) 
] ,
\end{equation}

\begin{equation}
B(r) = p_2 B_0 \rho ^2 (r) ,
\end{equation}

\begin{equation}
C(r) = p_2^{-1} C_0 \rho^2 (r) ,
\end{equation}

\begin{equation}
L(r) = \{ 1 + \frac{4 \pi}{3}  \lambda [ c(r) + C(r) ] \}^{-1} .
\end{equation}

\noindent
The kinematical factors are defined as;

\begin{equation}
p_1 = 1 + \omega / M  ,
\end{equation}

\begin{equation}
p_2 = 1 + \omega /2M  ,
\end{equation}

\noindent 
and the reduced energy $ \bar{\omega} $ is given by 

\begin{equation}
 \bar{\omega} = \omega / (1 + \omega / A M ) , 
\end{equation}

\noindent
where $\omega$ is the pion total energy, $\varepsilon_{\pi}$ is the pion charge, 
$A$ is the nuclear mass number, and $M$ is the nucleon mass.  
The potential is taken from Stricker $et$ $al.$ \cite{Stricker80}, which 
has been obtained by fitting to the pion nucleus elastic scattering cross 
sections.  We show the potential parameters in Table 1.  Needless to say, 
but the Coulomb potential is strongly attractive for the negative pions, 
while it is not present for the neutral pion.  The Klein-Gordon equation 
is solved by multipole expansion for all the necessary partial waves and 
the relevant pion energies.  

\section{Numerical Results on  $(n,d)$ reactions}

In this section we will show general features of the one nucleon 
pick-up pion transfer reactions by considering $(n,d)$ cases.  
We would like to first study the threshold behavior of the $(n,d)$ 
reactions leading to $\pi^-$ and $\pi^0$ states.  First of all, 
we show in Fig. 3 the pion wave functions for $l = 0$ at $T_\pi = 5 MeV$ 
obtained by solving the Klein-Gordon equation with and without the 
optical potential  
as an example.  It is important to note that the negative pion feels a 
strong Coulomb attraction, while the neutral pion does not.  This 
difference makes a large difference in the behaviors of the pion wave 
functions in the nuclear region as depicted in Fig. 3.  Without the 
optical potential, the neutral pion wave function is the spherical Bessel 
function $( j_0 )$.  The negative pion wave function is pulled in to the 
small $r$ region relative to the $\pi^0$ wave function as seen in Fig. 3 
(a).  By introduction of the optical potential, the central components of 
the wave functions are largely pushed outwards for both the negative and 
neutral pions as seen in Fig.3 (b).  The large difference remains in the 
nuclear interior region.  We comment here that these wave functions 
provide good elastic scattering cross sections.\cite{Stricker80}

These wave functions are then used in eq. (2) for $N_{eff}$ at various 
pion energies.  In order to calculate the energy spectra in the pion 
unbound region, we have to calculate cross sections for many final 
sates of the final nucleus $^{207}Pb$.  As a typical case, we show in 
Fig. 4 the quasi-elastic pion production cross sections for the cases of 
relatively large cross sections for $\pi^-$ and $\pi^0$ starting from the 
pion threshold energy.  Note that the hole states are neutrons for 
$\pi^-$ and protons for $\pi^0$.  This figure clearly shows the 
large difference between the threshold behaviors of negative and neutral 
pion formation cross sections.  The cross section for negative pion 
production at the threshold is finite, which smoothly connects to the 
finite cross sections in the bound pion region.  We make similar 
calculations for all the configurations 
of nucleon hole and pion particle states until the cross sections converge 
in the region of interest.  The single particle states included in the 
calculations are $1f_{7/2} - 3p_{1/2}$ neutron-states (16 states) and 
$1f_{7/2} - 3s_{1/2}$ proton-hole states (10 states).  

We show the calculated results in Fig. 5 leading to bound and unbound 
states.  In this figure, we assume the energy resolution to be 
$1 MeV$ $FWHM$ which is an actual value of the (n,d) experiments.  We 
will see, in the next section, more detail structure of the energy spectrum 
with better energy resolution which is realistic for the (d,$^3$He) reactions. 

The cross sections in the bound pion region are similar to those 
published before.\cite{Toki91}  In the unbound region, we find still a 
significant amount of strength coming from bound pionic atom states 
coupled with neutron holes at higher excitation energy; $s - d - g$ shell 
with $h_{11/2}$ intruder, which begins at $E_x \sim 9 MeV$ from the 
ground state of $^{207}Pb$.  On the other hand, the difference 
between the negative pion spectrum and the neutral pion spectrum is 
remarkable as can be seen by comparing the curves with $\pi^-$ and 
$\pi^0$.  The negative pion spectrum increases rapidly at the threshold, 
whereas the neutral pion spectrum only gradually increases from the 
threshold, which is $2.7 MeV$ below the negative pion threshold. The 
small jump seen at $8 MeV$ above the threshold in negative pion spectrum 
is caused by the beginning of the next neutron major shell with 
quasi-elastic pions.  Adding all the contributions, we get the total 
cross sections as depicted by the thick solid curve.  It is very 
interesting to note that the strength in the bound pion region is almost 
exclusively due to the bound pionic states.  The peak appearing at $B_\pi 
\sim 5 MeV$ mainly corresponds to the configurations 
$[l_\pi \otimes j_n^{-1}] = [2p \otimes p_{3/2}^{-1}]$ and 
$[2p \otimes p_{1/2}^{-1}]$ states.

Finally we overlay the calculated results on the experimental data, which 
is done in Fig. 6.  To do so, we multiply a normalization factor $N = 
1.7$ to the calculated results.  This factor seems a reasonable value as 
the uncertainty of the effective number approach.  We assume the flat 
background to be $0.28 [mb/sr/MeV]$ which is not due to the pionic 
contribution.  The agreement is remarkable.  A small peak structure seen 
in experiment in the bound pion region might be identified to the similar 
structures seen in the calculation.  

\section{Numerical Results on (d,$^3$He) reactions}

In this section, we compare our theoretical results with the data of the 
(d,$^3$He) reactions taken at GSI \cite{Yamazaki96,Toki96}.  We also 
show the spectrum with better energy resolution 
to predict what should be expected when experiments were performed with 
better energy resolution, and describe the 
underlying structures of the (d,$^3$He) excitation functions.  
We show the energy dependence of 
the excitation function of the (d,$^3$He) reaction with the $^{208}Pb$ 
target, too.    
For (d,$^3$He) reactions, we 
only need to change reaction kinematics, elementary cross section, 
and distortion cross sections used in the eikonal approximation for the 
projectile and the ejectile from $(n,d)$ cases described in section 2.  

We show the calculated results on $^{208}Pb (d,^3He)$ reactions at $T_d = 
600 MeV$ in Fig. 7.  In the upper part of Fig. 7, we show the forward 
cross sections with the resolution of 500 keV in order to compare with the 
experimental spectra shown in the middle part of Fig. 7.  The agreement 
is almost perfect.  We assume $ 20 [\mu b/sr/MeV]$ flat background in the 
theoretical spectra.  In order to see the underlying structures and also 
the spectra with a better experimental resolution, we show in the lower 
part of Fig. 7, the cross sections with 200 keV resolution.  We can see 
now many peaks below the pion production threshold, which is denoted by 
the vertical dashed line.  The peak seen experimentally consists of two 
states; $[l_\pi \otimes j_n^{-1}] = [p \otimes p_{1/2}^{-1}]$ and 
$[p \otimes p_{3/2}^{-1}]$.  Since 4 neutrons are allowed in the 
$p_{3/2}$ state and 2 neutrons in the $p_{1/2}$ state, the ratio of the 
cross sections leading to $p_{3/2}^{-1}$ and $p_{1/2}^{-1}$ is 2.  Those 
two states provide the peak and the shoulder structure at $Q= -136 MeV$ 
(around $5 MeV$ binding energy).  The pionic $s$ state seems to be only 
weakly excited due to the necessity of some momentum transfer.  In 
addition, there is an appreciable amount of contribution from the neutron 
$f_{5/2}$ hole coupled to the pionic $s$ state to this small peak.  
Many more peaks are seen closer to the pion production threshold.  The 
dominant contributions are again coming from the pionic atom states 
coupled with the $p_{3/2}$ and $p_{1/2}$ neutron holes.  It is 
interesting to comment on the bump structure seen in the calculated 
spectra around $Q=-143MeV$ in the continuum.  This bump structure is 
caused by exciting $[1s_\pi \otimes s_{1/2}^{-1}]$ state, where a 
$s_{1/2}$ neutron in the $50 < N < 82$ major shell is picked up and a 
pion is placed in the $1s$ orbit.  In the calculation, the neutron 
separation energy of the 
$s_{1/2}$ neutron hole state is calculated using the phenomenological 
potential \cite{Speth77} and width of the state is assumed to be zero 
since there is no experimental information.

We study now the energy dependence of the (d,$^3$He) cross sections.  As 
could be guessed from the energy dependence of the elementary 
differential cross sections  and that of momentum transfer, the 
(d,$^3$He) cross sections at different energies decrease.  It is, 
however, interesting to see how the cross sections will change with the 
incident energy.  For this purpose, we show the calculated results at 
$T_d = 500, 800,$ and $1000 MeV$ with the resolution of $\Delta E = 200 
keV$ and $500keV$ in Fig.8.  As shown in the figures, the energy resolution 
of the cross section is essentially important to see the structure of 
deeply bound pionic states.  

When we lower incident energy to $T_d = 500 MeV$, the peak structure at 
$E \sim 5MeV$ is enhanced relative to other structure due to smaller 
momentum transfer.  The peak structures are almost exclusively produced 
by the $p$ neutron hole contribution.  Note that the $1s$ pionic state is 
not appreciably excited at this energy.  Those energy dependence are 
clearly explained by using the energy dependence of the calculated 
effective numbers.\cite{Toki91,Hirenzaki91}  As the energy is increased to $T_d = 800 
MeV$, the cross sections are lowered.  As indicated by the three kinds of 
curves in the spectrum with $\Delta E = 200 
keV$, now the contributions from the $p$ neutron hole state become 
small, while the contributions from the $i_{13/2}$ and $f_{5/2}$ neutron 
hole states increase.  Pionic states coupled with $f_{5/2}$ neutron hole 
state are excited appreciably.  Further increased to $T_d = 1000 MeV$, 
the peak structures are completely dominated by the $i_{13/2}$ neutron 
hole state coupled with various pionic atom states.  

\section{Conclusion}

In this paper we have studied the $(n,d)$ and (d,$^3$He) reactions on 
$^{208}Pb$ for the formation of deeply bound pionic states.  We have 
calculated the formation cross sections for various cases and compared 
them with existing data.  Our model is found to describe the reaction 
very well and has good predictive power as proven by the fantastic 
agreement of newly obtained data to our prediction calculated before the 
experiments.  We believe the model is a clue for the systematic studies 
of the structure and formation of deeply bound pionic states.  

In order to develop the physics of deeply bound pionic atoms, 
we need to study the structure of pionic atoms more precisely by 
taking into account the configuration mixing effect of pion-particle 
neutron-hole state, the effect of nuclear deformation\cite{Nose97}, 
and other effects.  They may be important in deriving the pion mass in 
nuclear medium from 
the experiment\cite{Yamazaki97}.  We also need to study the formation 
cross sections to other target nuclei to look for proper cases for observing 
the certain states of deeply bound pionic states\cite{Hirenzaki97}.  It 
is also interesting to study the formation of the deeply bound pionic 
states on unstable nuclei\cite{Toki90} using the same reaction model.  
We also mention that this direct reaction can be applied to the formation 
of other mesons such as $\eta$ and $\omega$ inside nucleus to extract 
their properties at finite density\cite{Hirenzaki97-2}.  

\section{acknowledgment}
We acknowledge many discussions and collaborative works 
on the deeply bound pionic atoms with Prof. T. Yamazaki, Prof. R. S. Hayano and 
Dr. K. Itahashi.  We also acknowledge stimulating discussions 
with Prof. K. Kume.

\newpage

{\bf Table 1}
\\

Pion-nucleus optical potential parameters determined by Stricker 
$et$ $al.$\cite{Stricker80}
\\

\begin{center}
\begin{tabular}{cc} \hline
  $b_0 [fm]$ & -0.046 \\ 
  $b_1 [fm]$ & -0.134 \\
  $B_0 [fm^4]$ & 0.007 + 0.19$i$ \\
  $c_0 [fm^3]$ & 0.66 \\
  $c_1 [fm^3]$ & 0.428 \\
  $C_0 [fm^6]$ & 0.287 + 0.93$i$ \\
  $\lambda$ & 1.4 \\
  $k_F [fm^{-1}]$ & 1.4 \\  \hline
\end{tabular}
\end{center}

\newpage
{\bf Figure Caption}

\noindent
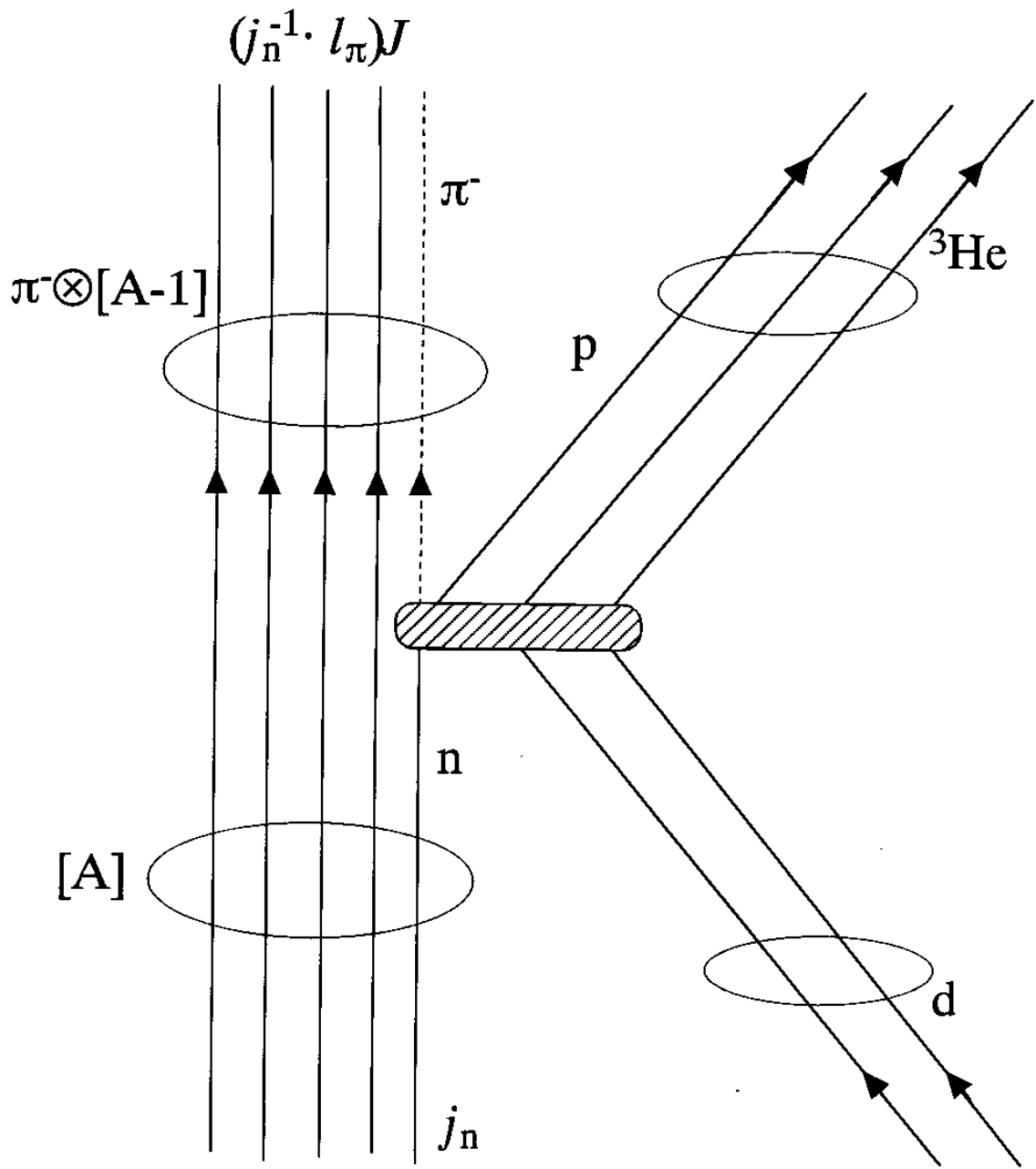
Fig. 1. Diagram for (d,$^3$He) reactions to form pionic bound states with a 
neutron-hole state.\\

\noindent
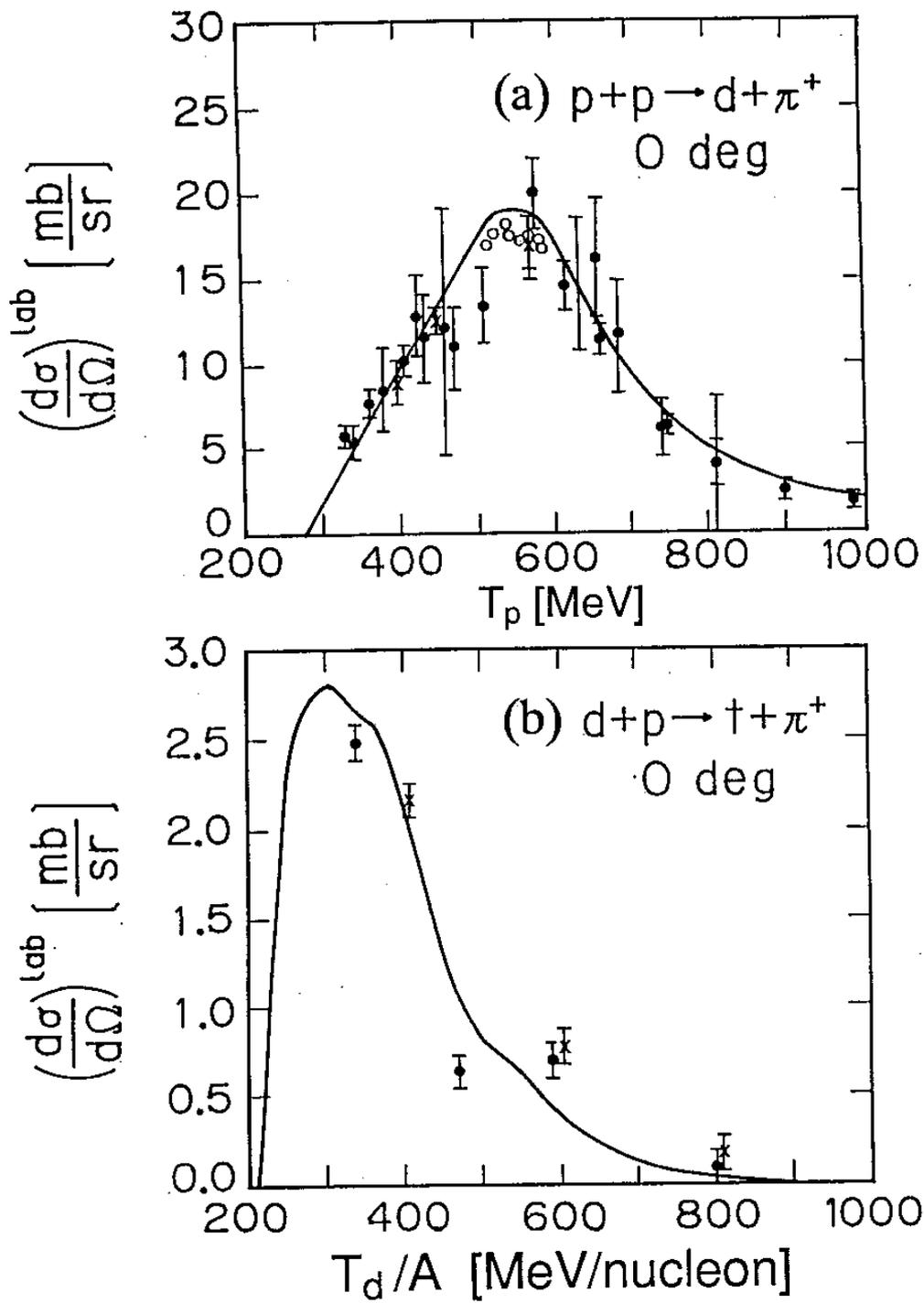
Fig. 2. (a) The differential cross section $(d\sigma/d\Omega)^{lab}$ at 
$0$ degree in the laboratory frame for $p(p,d)\pi^+$ as a function of 
the incident energy $T_p$, 
derived from the experimental values of $(d\sigma/d\Omega)^{c.m.}$ for 
$p(p,\pi^+)d$.\cite{Toki91}  The data are taken from references
\cite{Richard70,Aebischer76,Hoftiezer81}.  The solid line is the 
phenomenological fit to the data.\cite{Toki91}
(b) The differential cross section $(d\sigma/d\Omega)^{lab}$ at $0$ 
degree for $p(d,t)\pi^+$ as a function of the incident energy $T_d/A 
[MeV/nucleon]$, derived from the experimental values for $d(p,\pi^+)t$.
\cite{Fearing77,Aslanides77}  The solid curve is obtained by multiplying $1.7$ to the 
theoretical result of Fearing \cite{Fearing77} to match the experimental cross 
sections.\\

\noindent
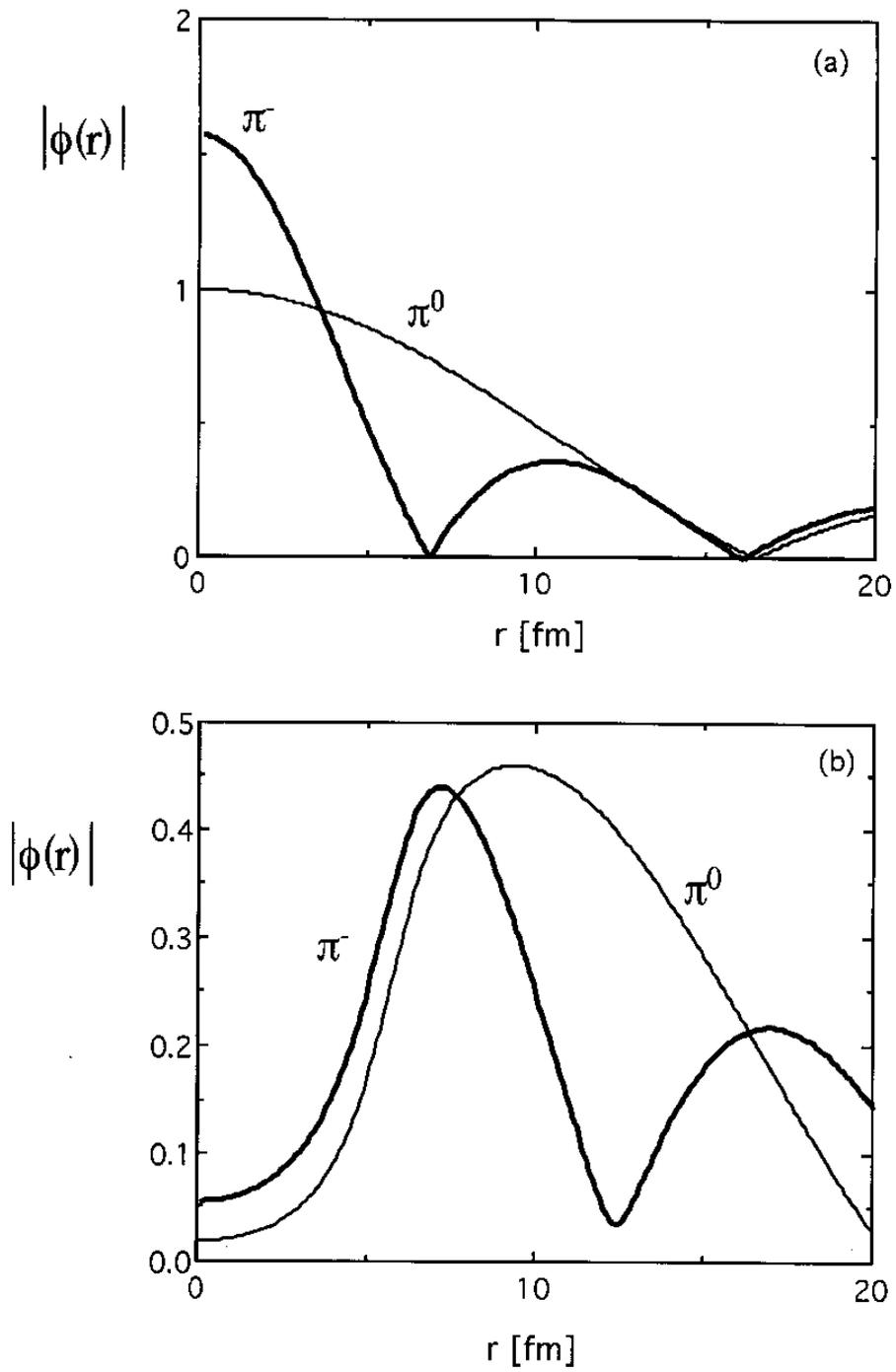
Fig. 3. The negative and neutral pion wave functions relative to 
$^{208}Pb$ at $T_\pi = 5 MeV$ for $l_{\pi}=0$.  (a) The wave functions without the 
optical potential.  (b) The wave functions calculated with optical 
potential, which reproduce the elastic pion scattering data.  \\

\noindent
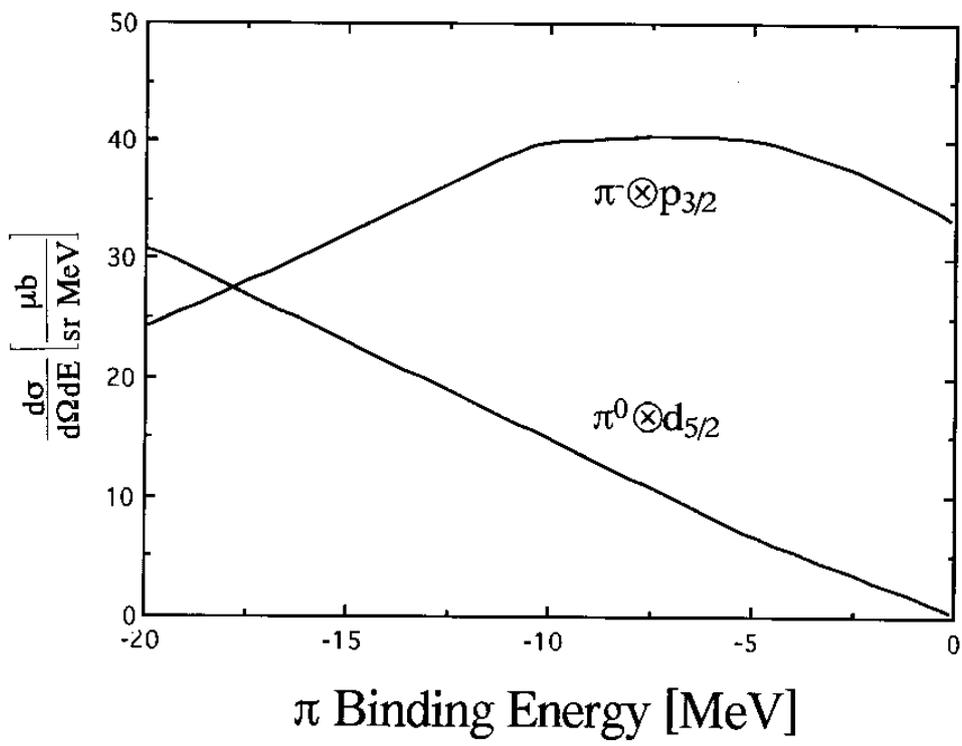
Fig. 4. Quasi-elastic pion production cross sections in $(n,d)$ reactions 
for a negative pion with $p_{3/2}$ neutron hole and those for a neutral 
pion with a $d_{5/2}$ proton hole.  The zero is the threshold energy for 
each case.  \\

\noindent
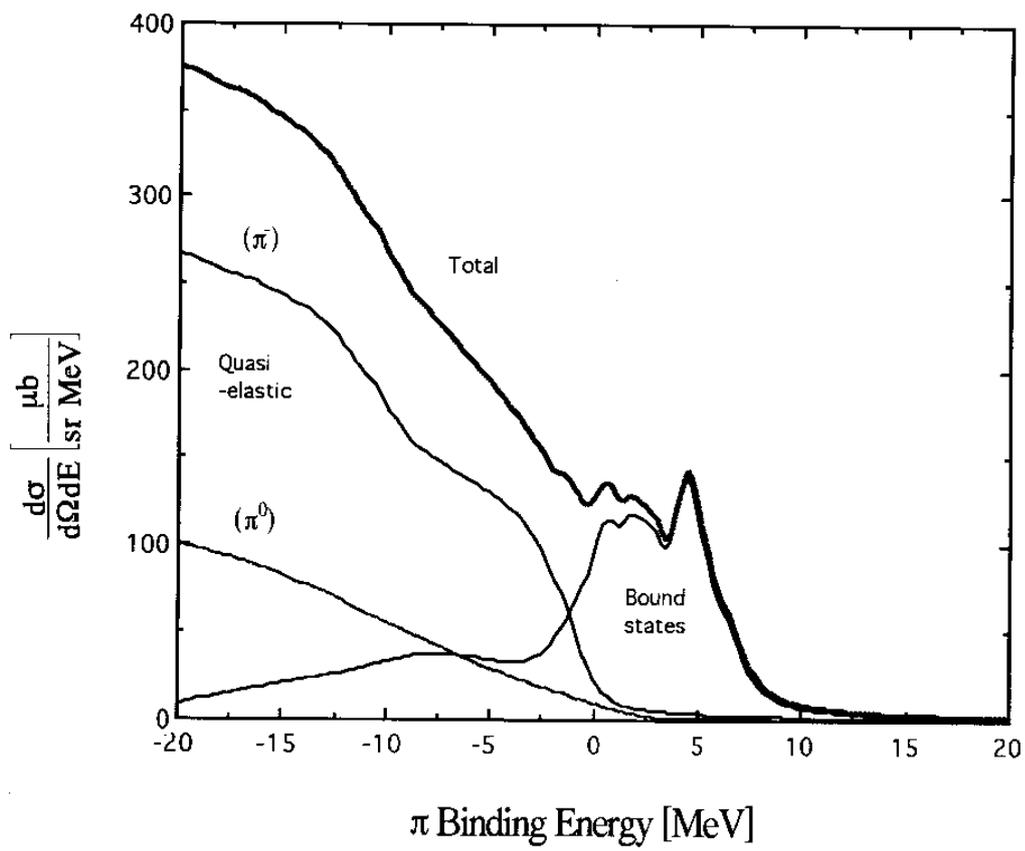
Fig. 5. $^{208}Pb$ reaction cross sections at zero degree leading to 
deeply bound pionic atoms and quasi-elastic pionic states calculated 
within the effective number approach.  The horizontal axis, 
$\pi$ Binding Energy, is 
defined to be $0 MeV$ at the $\pi^-$ production threshold with 
the ground state of $^{207}Pb$ and it 
corresponds to $Q = -143.4 [MeV]$ for $(n,d)$ reactions.  We take $1 MeV$ FWHM as an 
instrumental resolution.   \\

\noindent
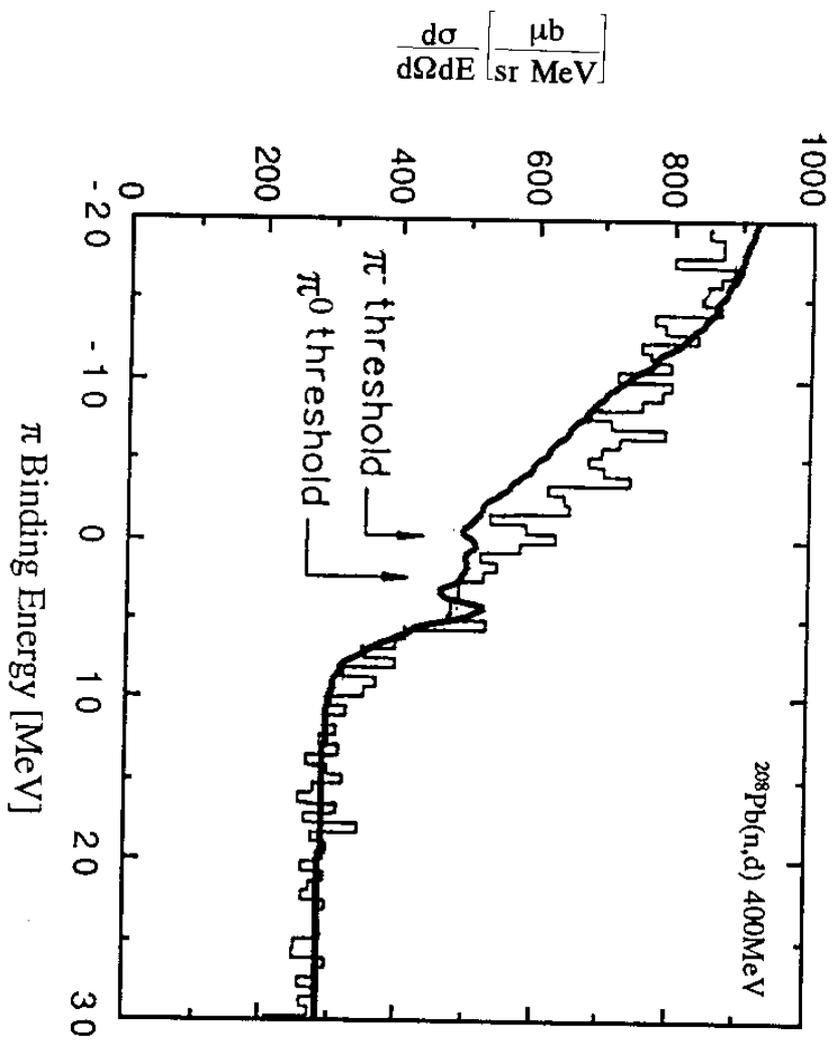
Fig. 6. $^{208}Pb (n,d)$ reaction cross sections at zero degree at $T_n = 
400 MeV$.  The solid curve is the calculated result with a normalization 
factor of $1.7$.  The flat background is assumed to be $280 [\mu 
b/sr/MeV]$.  We take $1MeV$ FWHM as an instrumental resolution.  The 
horizontal axis is defined as in Fig. 5.  \\

\noindent
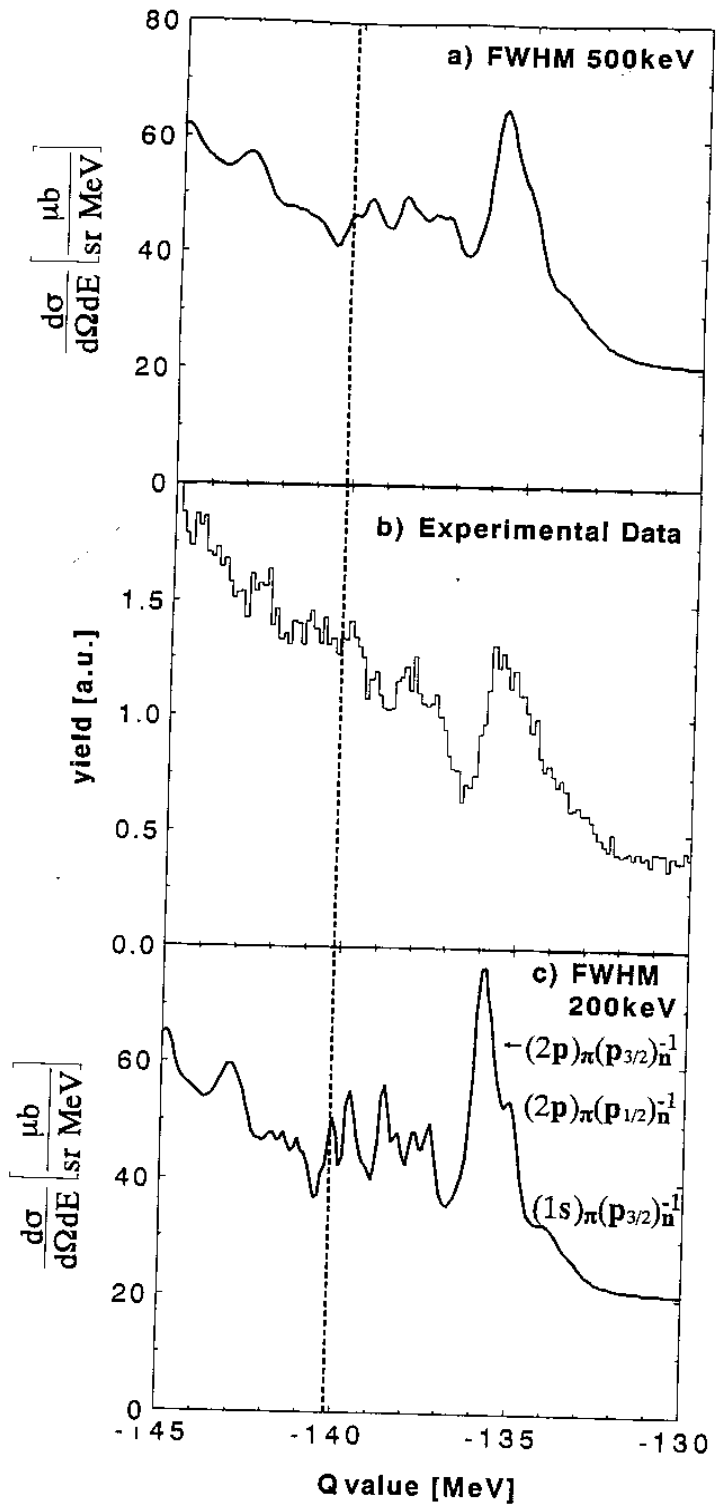
Fig. 7.  Forward cross sections of the $^{208}Pb$(d,$^3$He) reaction at 
$T_d = 600 MeV$; (a) Theoretical results with the experimental resolution 
of $ 500keV$, (b) Observed spectrum by Yamazaki $et$ $al.$ 
\cite{Yamazaki96}, and (c) Theoretical results with $200keV$ experimental 
resolution.  In (a) and (c) the flat background is assumed to be $ 20 [\mu 
b/sr/MeV]$.  The vertical dashed line denotes the $\pi^-$ emission 
threshold energy.  \\

\noindent
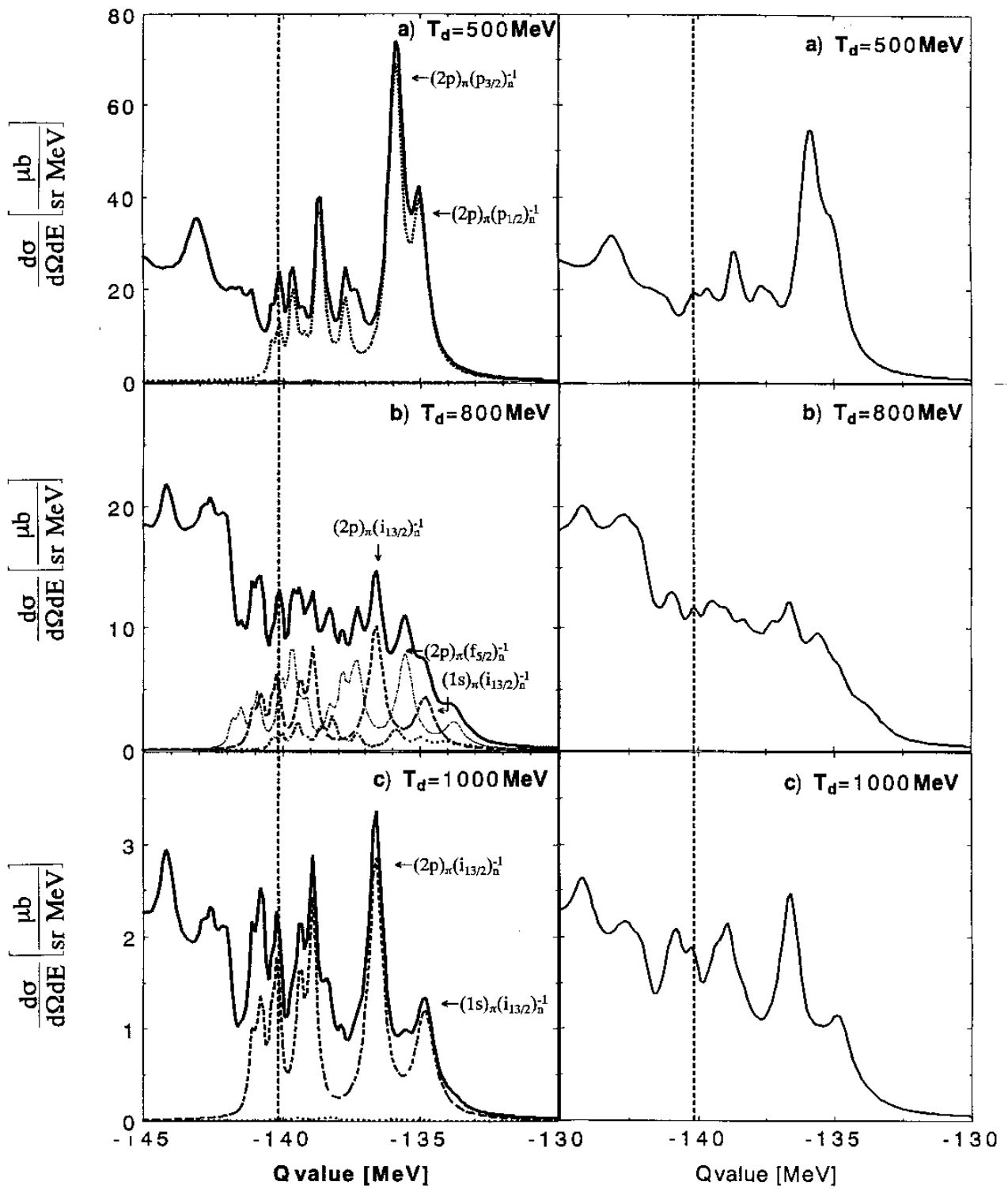
Fig. 8.  Calculated cross sections of the $^{208}Pb$(d,$^3$He) reaction at 
(a) $T_d = 500 MeV$, (b) $T_d = 800 MeV$, and (c) $T_d=1000 MeV$ with 
$200keV$ (left) and $500keV$ (right) experimental energy resolution.  
Solid curves show the full 
spectra including all contributions.  Dotted curves show the 
contributions from the neutron $2p$ states ($p_{3/2}$ and $p_{1/2}$) and 
dashed curve show those from the neutron $i_{13/2}$ state.  Added is also 
the contribution from the neutron $1f$ states ($f_{5/2}$ and $f_{7/2}$) 
by dash-dotted curve in the middle figure.  \\

\end{document}